**Objectively Evaluating the Reliability of Cell Type Annotation Using LLM-Based Strategies**

Wenjin Ye[1†], Junkai Xiang[2†], Yuanchen Ma[1,3†], Hongjie Liang[1], Tao Wang[1,4], Qiuling Xiang[1,5], Andy Peng Xiang[1,4,6], Weiqiang Li[1,4*], Weijun Huang[1,4*]

1. Center for Stem Cell Biology and Tissue Engineering, Key Laboratory for Stem Cells and Tissue Engineering, Ministry of Education, Sun Yat-Sen University, Guangzhou, Guangdong, 510080, China.

2. Department of Neurobiology, Physiology and Behavior, College of Biological Sciences, University of California, Davis, CA, 95616, USA.

3. Department of Gastrointestinal Surgery, The First Affiliated Hospital, Sun Yat-sen University, Guangzhou, Guangdong, 510080, China.

4. National-Local Joint Engineering Research Center for Stem Cells and Regenerative Medicine, Zhongshan School of Medicine, Sun Yat-Sen University, Guangzhou, Guangdong, 510080, China.

5. Department of Physiology, Zhongshan School of Medicine, Sun Yat-sen University, Guangzhou, Guangdong, 510080, China.

6. Department of Histoembryology and Cell Biology, Zhongshan School of Medicine, Sun Yat-Sen University, Guangzhou, Guangdong, 510080, China.

[†] Wenjin Ye, Junkai Xiang, Yuanchen Ma, contributed equally to this work.

***Corresponding authors:**

Weiqiang Li, Email: liweiq6@mail.sysu.edu.cn;

Weijun Huang, Email: hweijun@mail.sysu.edu.cn.

**Abstract**

Reliability in cell type annotation is challenging in single-cell RNA-sequencing data analysis because both expert-driven and automated methods can be biased or constrained by their training data, especially for novel or rare cell types. Although large language models (LLMs) are useful, our evaluation found that only a few matched expert annotations due to biased data sources and inflexible training inputs. To overcome these limitations, we developed the LICT (**L**arge language model-based **I**dentifier for **C**ell **T**ypes) software package using a multi-model fusion and "talk-to-machine" strategy. Tested across various single-cell RNA sequencing datasets, our approach significantly improved annotation reliability, especially in datasets with low cellular heterogeneity. Notably, we established objective criteria to assess annotation reliability using the "talk-to-machine" approach, which addresses discrepancies between our annotations and expert ones, enabling reliable evaluation even without reference data. This strategy enhances annotation credibility and sets the stage for advancing future LLM-based cell type annotation methods.

**Key words:** Large language model, Cell type annotation, Annotation reliability, Multi-model fusion, Talk-to-machine

# Introduction

Cell type annotation is crucial for understanding cellular composition and function in single-cell RNA sequencing (scRNA-seq) data, making it an indispensable step in data exploration[1]. Traditionally, this annotation process has been performed either manually or with automated tools. Manual annotation benefits from expert knowledge but is inherently subjective and highly dependent on the annotator's experience. On the other hand, automated tools provide greater objectivity but often depend on reference datasets, which can limit their accuracy and generalizability (Figure 1a)[2-4].

Recent advancements in artificial intelligence (AI) have opened new possibilities for cell type annotation. One promising development is GPTCelltype, a tool leveraging the large language model (LLM) ChatGPT, which demonstrates that LLMs can autonomously perform cell type annotations without requiring extensive domain expertise or reference datasets[1]. However, since LLMs are not specifically designed for cell type annotation and are trained on diverse data sources, only a few are well-suited for this task. Even among these, no single model can accurately annotate all cell types. Moreover, the standardized data format encoded within LLMs limits their ability to adapt to the dynamic and complex nature of biological data, where different pieces of evidence may converge on the same conclusion or diverge[5]. Thus, new strategies are needed to enhance LLMs' adaptability and improve their performance in cell type annotation, potentially through advanced self-learning mechanisms and ongoing model updates[6].

To address the limitations of LLM-based cell type annotation, we conducted a comprehensive comparison of available models to identify those that excel in cell annotation tasks. Building on this, we developed the LICT (**L**arge language model-based **I**dentifier for **C**ell **T**ypes) software package, which incorporates the most effective LLMs. LICT employs three distinct strategies to enhance the credibility of annotation results. First, a multi-model fusion strategy combines the strengths of various LLMs, reducing uncertainty and ensuring reliable outputs. Second, a "talk-to-machine" strategy iteratively provides LLMs with additional context and information, significantly mitigating ambiguous or biased annotations. Third, an objective credibility evaluation strategy leverages the data itself as a benchmark for assessing annotation credibility, ensuring objectivity, especially when accurate reference data is unavailable (Figure 1b).

Together, LICT, with its optimized strategies, harnesses the full potential of large language models to enhance cell type identification, improving both the reliability and robustness of cell type determinations. These strategies not only overcome current limitations of cell annotation but also pave the way for future advancements in automated, AI-driven biological data analysis.

**Results**

**Identification of top-performing LLMs for cell type annotation**

To identify the most effective large language models (LLMs) for cell type annotation,

we conducted an initial evaluation of 87 different LLMs[7] using a scRNA-seq dataset of PBMCs (GSE164378)[8]. Each model was tested with standardized prompts to ensure consistency in the evaluation process, and their annotation outputs were systematically recorded (Table S1). Based on accessibility and annotation accuracy, we selected five top-performing LLMs for further analysis: GPT-4[9], LLaMA-3[10], Claude 3[11], Gemini[12], and the Chinese language model ERNIE 4.0[13] (Table 1). These models were then integrated into our LICT software package.

**Performance of LLMs diminishes when annotating less heterogeneity datasets**

We began by assessing the credibility of cell annotations across five LLMs. The four scRNA-seq datasets used in this study represented diverse biological contexts, including normal conditions (e.g., peripheral blood mononuclear cells or PBMCs[8]), developmental stages (e.g., human embryos[14]), disease states (e.g., gastric cancer[15]), and low-heterogeneity cellular environments (e.g., stromal cells in mouse organs)[16]. For each dataset, we selected the top 10 differentially expressed genes and applied a basic prompting strategy to generate cell type annotations. The accuracy of these annotations was then evaluated using the numerical scoring methodology proposed by Wenpin Hou and Zhicheng Ji[1].

First, we evaluate each selected LLMs with 4 independent datasets (figure 2a). The results showed that all selected LLMs excelled in annotating highly heterogeneous cell subpopulations, such as those in PBMCs and gastric cancer samples, with Claude 3 demonstrating the highest overall performance (Figure 2b-c). However, significant discrepancies emerged when annotating less heterogeneous subpopulations, such as those in human embryos and stromal cells, compared to manual annotations. Among the top-performing models, Gemini 1.5 Pro achieved 39.4% consistency with manual annotations for embryo data, while ERNIE 4.0 reached 29.2% consistency for fibroblast data (Figure 2d-e). The annotation consistent average score was shown in Figure 2f. These findings highlight the importance of integrating multiple LLMs to achieve more comprehensive and reliable cell annotations[17, 18].

**Strategy I: multi-model fusion strategy**

To improve LLM performance, particularly for low-heterogeneity datasets, we developed a multi-model fusion strategy. This approach integrates the annotation outputs from all five LLMs, selecting the result that most closely aligns with manual annotations for each cell type (Figure 3a). This strategy significantly reduced the mismatch rate in highly heterogeneous datasets—from 21.5% to 9.7% for PBMCs and from 11.1% to 8.3% for gastric cancer data—compared to GPTCelltype. Across all datasets, over 90% of our annotations were consistent with manual annotations, including both complete and partial matches (Figure 3b-c). For low-heterogeneity datasets, the improvement was even more pronounced, with alignment rates increased to 30% for embryo and 12.5% fibroblast data (Figure 3d-e). Despite these gains, discrepancies remain, with over 50% of annotations for low-heterogeneity cells still not matching manual results. The annotation consistent average score was shown in Figure

3f. This suggests that richer informational contexts in high-heterogeneity data may contribute to more robust model training and annotation accuracy, while also highlighting ongoing opportunities for refining the annotation process[19].

**Strategy II: "talk-to-machine" strategy**

To address the limitations of LLM credibility in annotating low-heterogeneity cell types, we implemented a "talk-to-machine" strategy to enhance annotation precision (Figure 4a). This human-computer interaction process involves the following steps: (1) Directing LICT to generate characteristic genes for each annotated cell type. (2) Analyzing the input single-cell RNA sequencing data to assess the expression of these characteristic genes within the corresponding cell types. (3) Confirming accurate cell annotation if the expression of more than four characteristic genes aligns with expected patterns. (4) If this criterion is not met, LICT is notified, and the cell annotation process is restarted, characteristics genes combine with additional DEGs from original single cell data will provide and ask LLM to update its response. This approach is designed to optimize the precision of cell annotations.

Under this optimization strategy, the alignment between our annotation results and manual annotations has significantly improved. In highly heterogeneous cell datasets, the rate of full match reached to 34.4% for PBMC and 69.4% for gastric cancer, with mismatch reduced to 7.5% and 2.8%, respectively (Figure 4b-c). Similarly, in low-heterogeneity cell datasets, full match rate improved by 16-fold for embryo data compare to simply use GPT-4, reaching 48.5%, while the full match rate for fibroblast data reached 37.5%, with mismatch decreasing to 42.4% and remain 62.5% (Figure 4d-e). The annotation consistent average score was shown in Figure 4f. These results demonstrate that our interactive LLM strategy successfully enhances annotation accuracy for both high- and low-heterogeneity datasets[20-22]. However, effort still needed since there still remains over 50% inconsistency of annotations in low-heterogeneity data.

**Strategy III: objective credibility evaluation**

Despite approximately 50% of low-heterogeneity cell type annotations failing to align with manual results, it is important to recognize that discrepancies between LLM and manual annotations do not necessarily indicate lower reliability of LLM results. Manual annotations can vary and be biased, particularly in low-heterogeneity datasets where cell clusters may be ambiguous[23-25]. Without a universally objective standard for evaluation, assessing the credibility of annotations remains challenging[26-28]. Our "talk-to-machine" strategy provides an objective framework for evaluating discrepancies by leveraging a data-driven approach, enabling more informed judgments about annotation reliability[29, 30] (Figure 5a).

The results of our credibility evaluation showed that the reliability of annotations was comparable between expert manual annotations and those generated by LICT in high-heterogeneity datasets (Figure 5b-c). In low-heterogeneity datasets, LICT annotations outperformed expert ones. Specifically, LICT annotations were deemed credible in 50% of mismatch cases for the embryo dataset, compared to just 21.3% for expert annotations. For stromal cell data, the figures were even more pronounced, with 42.9% of LICT annotations considered credible versus 0 for expert annotations (Figure 5d-e). Additionally, our objective evaluation highlighted cases where discrepancies between LICT and expert annotations existed, yet both were considered reliable, accounting for approximately 14.3% of the embryonic dataset. These findings provide valuable insights into the underlying cellular properties and states from diverse perspectives, enhancing our ability to interpret complex biological data.

**Generalizability of our optimization strategy**

To assess the generalizability of our annotation strategy, we applied it to two freely available LLMs, LLaMA-3 and Gemini, using the same optimization techniques as those applied to LICT (Figure 6a). This approach significantly enhanced annotation reliability, with consistency with manual annotations improving by 5.5% to 15.2% (Figure 6b-e). While the performance of these two models individually did not match the results obtained with the five-model integration, they still outperformed single models like GPT-4, demonstrating the effectiveness of our strategy. These findings underscore the importance of both the number and quality of LLMs in achieving reliable cell annotations.

**Discussion**

In summary, to bolster the credibility of cell annotation results for large language models, we employed a multi-model fusion strategy and a feedback-driven human-machine interaction approach. Objective evaluation criteria were established to gauge the reliability of annotation outcomes. Validated across diverse single-cell RNA sequencing datasets, our integrated optimization strategy significantly enhanced the credibility of cell annotation results, particularly for datasets with low cellular heterogeneity. Furthermore, in instances of discrepancies between our annotations and expert annotations, our strategy facilitated an objective determination of which result was more reliable, even at no reference available situation.

While LLMs can be useful for cell annotation, our evaluation found that only a few closely matched expert annotations. Even these high-performing models did not consistently deliver reliable results across all cell types[31-34], as each LLM excels in different areas[35]. For example, GPT-4 performed well on PBMC and gastric cancer datasets but was less effective on human embryo data. The Chinese LLM ERNIE 4.0 was most effective for low-heterogeneity datasets, such as stromal cell data. These variations in performance likely arise from differences in data sources, the amount of training data used, and the rigid input formats employed during LLM training, which can distort or lose information in complex, highly variable biological data. Our tests also showed that LLMs are less consistent with low-heterogeneity data, possibly due to information degradation caused by rigid input formats. Based on these findings, relying on a single LLM for cell annotation is unlikely to be optimal. Instead, we propose a multi-model fusion strategy, selecting the five LLMs that currently perform best in cell annotation. By integrating their results, we can achieve more accurate and reliable annotations. This approach capitalizes the strengths of each model and has proven effective, increasing the match rate between our annotations and expert annotations by at least 2.8% for gastric cancer data and up to 15.2% for human embryo data.

Given the training methods used for LLMs, it's important to acknowledge that LLMs are not specifically designed for cell type annotation. To fully capitalize on the benefits of a model fusion strategy, a number of specialized models tailored for cell annotation are necessary. While there remains scFoundation serves as an example of such a model[36], the limited availability of these models restricts the potential for comprehensive cell annotation, thereby limiting the optimization of this approach. Additionally, cell annotation models face challenges similar to those of LLMs, such as biased data sources and inflexible training inputs. As a result, relying solely on a model fusion strategy does not fully address the challenges of cell annotation reliability.

To overcome these limitations, we developed a "talk-to-machine" strategy that equips LLMs with self-correction capabilities, reducing ambiguous or biased annotations. By providing LLMs with additional context and information, this approach enables LLMs to objectively assess the credibility of annotation results and make necessary revisions to those deemed unreliable, effectively minimizing discrepancies between LLM-generated and expert annotations. For example, when this strategy is combined with multi-model fusion, the fully match rate between annotations and expert results increases by at least 12.5% for fibroblast data and up to 45.5% for human embryo data compared to using GPT-4 alone. This suggests that LLMs are beginning to process and utilize the additional information in a manner similar to human inference. Interestingly, we found that current LLMs achieve optimal results when this strategy is applied only once; repeated applications do not significantly enhance the credibility of the annotation outcomes. This diminishing effectiveness with repeated inquiries is a phenomenon commonly observed in other LLMs applications and may be related to the inherent limitations of the LLMs architecture and their interaction styles.

Building on the "talk-to-machine" strategy, we developed an objective evaluation approach to address the challenges of assessing annotation credibility, especially in the absence of a universally accepted standard. This approach leverages the data itself to objectively reflect its characteristics, thereby avoiding the pitfalls of subjective interpretation[37]. It ensures that LLM outputs are rational, with each response being accountable and verifiable. Through this method, we found that discrepancies between LLMs and manual annotations do not necessarily suggest that LLM results are less reliable. For instance, LLM-generated mismatches in PBMC, human embryo and fibroblast datasets were actually more reliable than expert annotations. It was a reasonable outcome, given that sole reliance on expert judgment can introduce subjective bias. Additionally, our objective evaluation identified instances where inconsistencies between LLM and expert annotations were present, yet both were deemed credible, offering valuable insights that can enhance our understanding of cellular properties and states from multiple perspectives.

The objective evaluation strategy holds substantial potential as a robust selection criterion, ensuring consistent and unbiased results regardless of the evaluator's expertise. By assessing all available cell annotation data sources without introducing subjective bias, we can identify the most reliable annotations. Leveraging this high-quality data to train LLMs mitigates the limitations of rigid training data formats, thereby enhancing the accuracy and effectiveness of cell annotation by the models.

Overall, our LLM optimization strategies have proven effective in enhancing concordance with manual annotations and delivering reliable results even when references are unavailable. These strategies not only improve annotation credibility and objectivity but also contribute to refining future LLM-based cell type annotation efforts. They lay the groundwork for further advancements in automated, AI-driven biological data analysis.

## Online Methods

### Dataset collection

Four datasets used in this study: human peripheral blood mononuclear cells (PBMCs, GSE164378, https://www.ncbi.nlm.nih.gov/geo/query/acc.cgi?acc=GSE164378)[8], fibroblasts from various organs(https://fibroXplorer.com.)[16], gastric tumor

samples(GSE206785, http://ncbi.nlm.nih.gov/geo/query/acc.cgi?acc=GSE206785)[15], and human embryo data(http://www.human-gastrula.net)[14]. These datasets were chosen based on their diverse cell types and comprehensive annotations. PBMCs and gastric cancer datasets are characterized by their heterogeneity, reflecting a wide range of different cell types, while the embryo and fibroblast datasets are highly homogeneous, representing more uniform cell populations. This selection allows us to evaluate whether there are strategies to optimize the performance of large language models (LLMs) for both homogeneous and heterogeneous datasets. Manually annotated cell type labels from each dataset were also collected for validation purposes.

**Dataset Processing**

For each dataset, we performed data processing steps such as quality control, dimensionality reduction, and clustering, following the protocols outlined in the respective original articles. Differentially expressed gene (DEG) calculations were conducted for each cell cluster using the *FindAllMarkers* function in the Seurat package (version 4.3.0). We selected genes with a $\log_2$ fold change greater than 0.5 and an expression percentage above 25% for further analysis. These genes were ranked in descending order based on their $\log_2$ fold changes. In cases where multiple genes had the same average $\log_2$ fold change, they were further ranked by their *p*-values.

**Model and API selection**

Five large language models (LLMs) were selected for cell type annotation based on their accessibility and efficiency: GPT-4 Turbo Preview (June 2024 version), Gemini 1.5 Pro (June 2024 version), Claude 3 Opus (June 2024 version), Llama 3 70B (June 2024 version), and ERNIE-4.0-8K (June 2024 version). We chose these models due to their advanced natural language processing capabilities and their robust APIs, which facilitate seamless integration with the dataset annotation tasks.

**Assessing LLMs cell annotations**

Cell type annotations provided by the LLMs were compared directly to the manual annotations from the original studies. To enhance objectivity, we introduced cell ontology (CL) names to each manually or automatically identified cell type annotation. Pairs of manually and automatically identified cell annotations were classified as 'fully match' if they shared the same annotation term or available CL cell ontology name, 'partially match' if they shared a broad cell type name or subordinate relationship (e.g., fibroblast and stromal cell), but had different annotations and CL names, and 'mismatch' if they had different broad cell type names, annotations, and CL names.

We assigned agreement scores of 1, 0.5, and 0 for 'fully match', 'partially match', and 'mismatch', respectively. Average scores were calculated for each dataset across cell types and tissues to facilitate comparison.

**Prompt for cell type annotation**

To efficiently transfer single-cell DEGs to each LLM, we developed an R package called LICT (**L**arge language model-based **I**dentifier for **C**ell **T**ypes). This package received DEG information calculated from the Seurat package and transfer them into a formatted prompt. Then five independent functions were built in order to send these formatted prompts to each LLMs.

Three types of prompts were constructed based on the operating environment:

'Identify cell types of species tissuename using the following markers. Identify one cell type for each row. Just reply with the cell type; there is no need to reply to the reasoning section or explanation section. \n GeneList \n Reply in the following format: \n 1: xx \n 2: xx \n N: xx \n

N is the line number, xx is a phrase including only cell types, such as pluripotent stem cells and smooth muscle cells.'

In this prompt, "(\n)" represents a newline character. The placeholder's species and 'tissuename' are replaced with actual species (e.g., human, mouse) and tissue names (e.g., brain). 'GeneList' contains the differential genes, with genes for each cell population separated by commas, and different populations separated by newline characters.

**Prompt for Validation of Cell Annotations:**

'Provide 15 marker genes for CellType, separated by ', ' between each gene.'

Here, 'CellType' refers to the annotated cell type results from the LLMs. We query GPT-

4 with this prompt to identify the genes that should be expressed for the specified cell type. The response is then input into the *Validate()* function of the LICT package to verify the gene expression in the data. Cell type annotations with four or more positive marker genes (expressed in over 60% of cells) are considered validated. Characteristic genes will be labeled as "Positive Gene" if expressed in over 60% of cells, or as "Negative Gene" if expressed in 60% or fewer cells.

**Prompt for Refining Responses:**

'Positive GeneList is expressed in the %d row\n Negative GeneList is expressed in the %d row \nBased on the additional information above, modify my previous response and list all cell types, including those that have not been modified. Reply in the following format: \n1: xx \n2: xx \nN: xx \nN is the line number, xx is the cell type.'

If the cell type annotation fails validation, an additional prompt is triggered. This prompt provides: (1) characteristic genes with their corresponding labels ("Positive Gene" or "Negative Gene") and (2) differentially expressed genes (DEGs) from the original data, ranked 11th to 20th by $\log_2$ fold change. These inputs are then used to request that the LLMs refine their previous response.

**Annotation evaluation framework**

Our "talk-to-machine" strategy offers an objective framework for assessing annotation reliability. The process includes the following steps: (1) Request LICT to provide characteristic genes based on the annotation results, (2) examine the expression of these genes in the original scRNA-seq data to determine which are definitively expressed, and (3) if the number of expressed characteristic genes exceeds the pre-set threshold, the annotation is deemed reliable; otherwise, it is considered unreliable. All evaluations are based on these uniformed criteria, with information derived solely from the LLM and the original dataset, ensuring objective, evidence-based conclusions.

**Availability of data and code**

The LICT package (v0.1.0) is provided as an open-source software package with a detailed user manual available in the GitHub repository at https://github.com/Glowworm-cell/LICT. All codes to reproduce the presented analyses are publicly available in the GitHub repository at https://github.com/Glowworm-cell/LICT_paper.

**Competing interests**

All authors declare they have no conflicting interest.

**Authors' contributions**

WY, YM and WH conceptualized and designed the project;

APX, WL and WH supervised the research;

JX performed data collection;

WY, JX and HL performed experiments and data analysis;

WJ, JX and HL designed the LICT packages;

MY and WH drafted the manuscript;

APX, WL, YM and WH revised the manuscript;

All authors read and approved the final manuscript.

**Acknowledgments**

This work was supported by grants from the National Key Research and Development Program of China (2021YFA1100600, 2023YFC2506100, 2022YFA1104100); the National Natural Science Foundation of China (82270566, 82471462, 32130046); Science and Technology Program of Guangzhou (202206060003); Pioneering talents project of Guangzhou Development Zone (2017-L163).

**Table 1. The 5 large language models most closely aligned with expert annotations**

| Model | Company | Website | Number of cell type | Response number | Free | API | Match | Mismatch |
|---|---|---|---|---|---|---|---|---|
| **Claude 3 opus** | Anthropic | claude.ai | 31 | 31 | No | Yes | 26 | 5 |
| **Llama 3 70B** | Meta | llama.meta.com | 31 | 31 | No | No | 25 | 6 |
| **ERNIE-4.0** | Baidu | qianfan.cloud.baidu.com | 31 | 31 | No | Yes | 25 | 6 |
| **GPT4** | OpenAI | openai.com | 31 | 31 | No | Yes | 24 | 7 |
| **Gemini 1.5 pro** | Google | deepmind.google/technologies/gemini/pro | 31 | 31 | Yes | Yes | 24 | 7 |

**Figure legends**

**Figure 1. Comparison of current cell type annotation methods.**

**A.** Comparison of expert-driven, automated, and ChatGPT-based methods for cell type annotation. Each column represents a specific annotation aspect, with the number of stars indicating the level of susceptibility, ranging from most (three stars) to least (one star)

**B.** Schematic diagram of the LICT software analysis process. Arrows indicate the processing sequence, which includes data preparation, consultation, validation, and feedback.

**Figure 2. Annotation performance of the top five LLMs across four independent datasets.**

**A.** Workflow of the entire study. The light green square highlights the current analysis process.

**B-E,** Bar plot illustrating the performance of each LLM across PBMC, gastric cancer, embryo, and fibroblast datasets. The x-axis indicates the match percentage between LLM and expert annotations, while the y-axis lists the LLMs. Bar colors represent match categories: full match (dark green), partial match (light green), and mismatch (grey).

**F.** Heatmap summarizing match rates for each LLMs. The x-axis represents the LLMs, while the y-axis corresponds to four independent datasets. Colors range from red (high match rate score) to light red (low match rate score), providing a visual representation of model performance across datasets.

**Figure 3. The multi-model fusion strategy improves the annotation performance of LLM.**

**A.** Workflow of the entire study. The light green square highlights the current analysis process.

**B-E,** Bar plot illustrating the performance of ChatGPT-4 and the integrated set of five LLMs across PBMC, gastric cancer, embryo, and fibroblast datasets. The x-axis shows the match percentage between LLM and expert annotations, while the y-axis, from top to bottom, represents the performance of ChatGPT-4 and the integrated performance of five LLMs. Bar colors represent match categories: full match (dark green), partial match (light green), and mismatch (grey).

**F.** Heatmap summarizing match rates for ChatGPT-4 and the integrated set of five LLMs. The x-axis represents the performance of ChatGPT-4 and the integrated performance of five LLMs, while the y-axis corresponds to four independent datasets. Colors range from red (high match rate score) to light red (low match rate score), providing a visual representation of model performance across datasets.

**Figure 4. Integrating both the multi-model fusion and 'talk-to-machine' strategies further enhances the annotation performance of LLM.**

**A.** Workflow of the entire study. The light green square highlights the current analysis process.

**B-E,** Bar plot illustrating the performance of ChatGPT-4, the multi-model fusion strategy alone, and the integration of both strategies across PBMC, gastric cancer, embryo, and fibroblast datasets. The x-axis shows the match percentage between LLM and expert annotations, while the y-axis, from top to bottom, represents the performance of ChatGPT-4, the multi-model fusion strategy alone, and the integration of both strategies. Bar colors represent match categories: full match (dark green), partial match (light green), and mismatch (grey).

**F.** Heatmap summarizing match rates for ChatGPT-4, the multi-model fusion strategy alone, and the integration of both strategies. The x-axis represents the performance of ChatGPT-4, the multi-model fusion strategy alone, and the integration of both strategies, while the y-axis corresponds to four independent datasets. Colors range from red (high match rate score) to light red (low match rate score), providing a visual representation of model performance across datasets.

**Figure 5. Reliability of LLM and expert annotations within mismatch results.**

**A.** Workflow of the entire study. The light green square highlights the current analysis process.

**B-E,** The top bar plot illustrates the performance of integrating both strategies across PBMC, gastric cancer, embryo, and fibroblast datasets. The bottom bar plot shows the reliability distribution of LLM and expert annotations. Bar colors denote match categories: full match (dark green), partial match (light green), and mismatch (grey). Reliability categories are indicated as follows: both reliable (dark blue), only LLM reliable (light blue), only expert reliable (dark yellow), and both unreliable (light yellow).

**Figure 6. Generalizability of our optimization strategy.**

**A.** Workflow of the entire study. The light green square highlights the current analysis process.

**B-E,** The left bar plot illustrates the performance of ChatGPT-4, the integration of both strategies using free LLMs, and all five LLMs across PBMC, gastric cancer, embryo, and fibroblast datasets. The x-axis represents the match percentage between LLM and expert annotations, while the y-axis, from top to bottom, shows the performance of ChatGPT-4, the integration of both strategies using free LLMs, and all five LLMs. The right bar plot depicts the reliability distribution of LLM and expert annotations within mismatch results. Bar colors denote match categories: full match (dark green), partial match (light green), and mismatch (grey). Reliability categories are indicated as follows: both reliable (dark blue), only LLM reliable (light blue), only expert reliable (dark yellow), and both unreliable (light yellow).

# Figures

## Figure 1. Comparison of current cell type annotation methods.

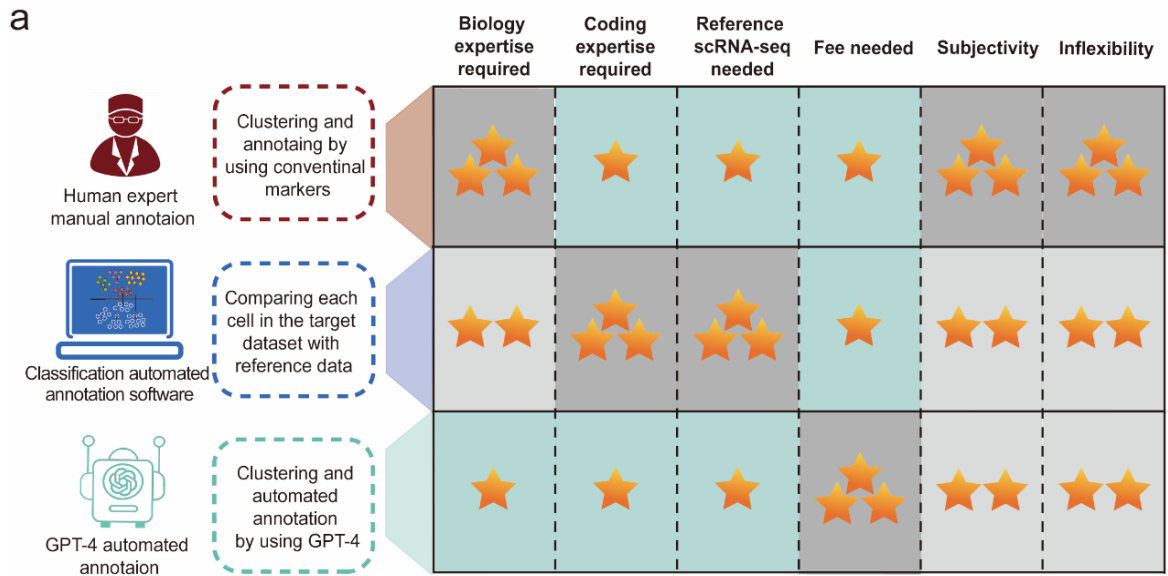

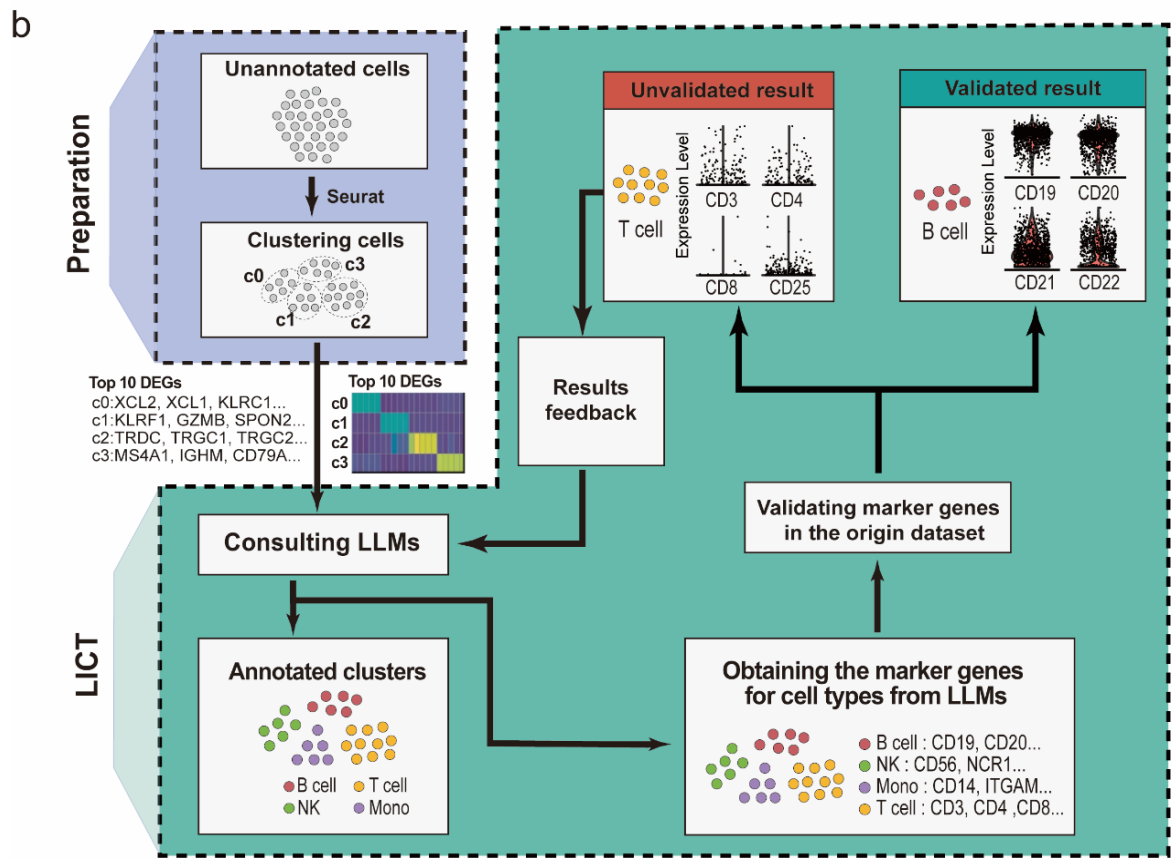

**Figure 2. Annotation performance of the top five LLMs across four independent datasets.**

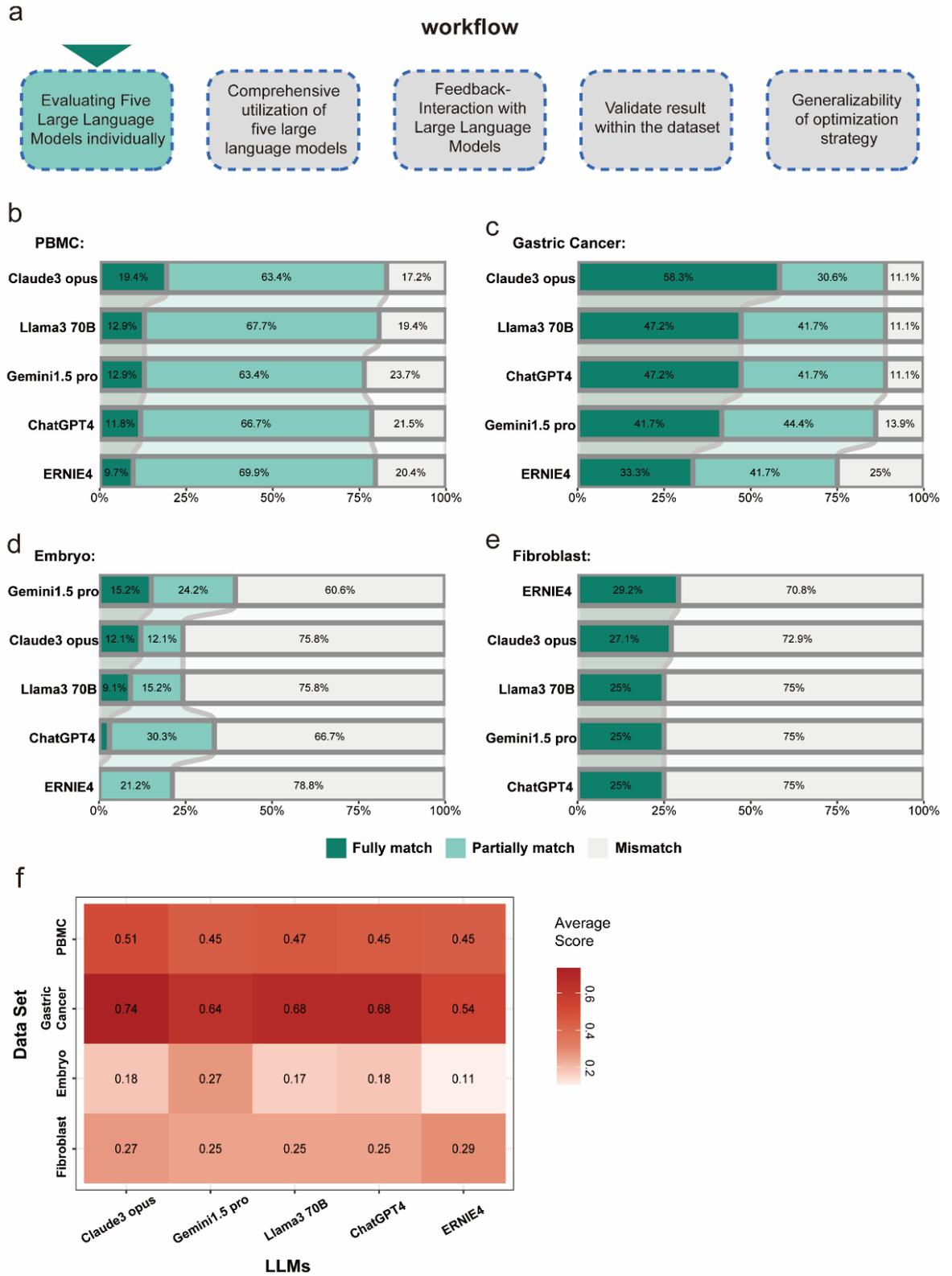

**Figure 3. The multi-model fusion strategy improves the annotation performance of LLM.**

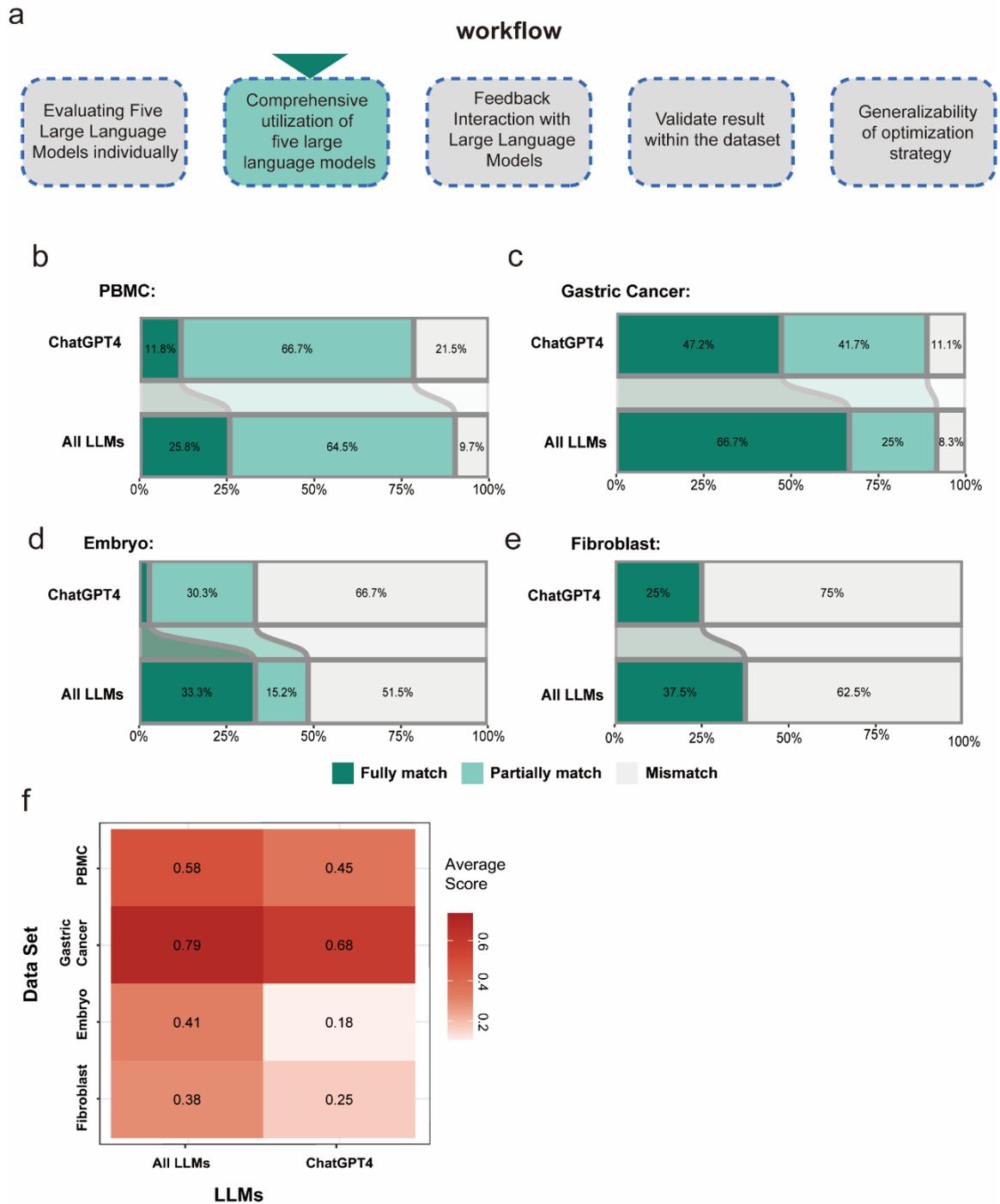

**Figure 4. Integrating both the multi-model fusion and 'talk-to-machine' strategies further enhances the annotation performance of LLM.**

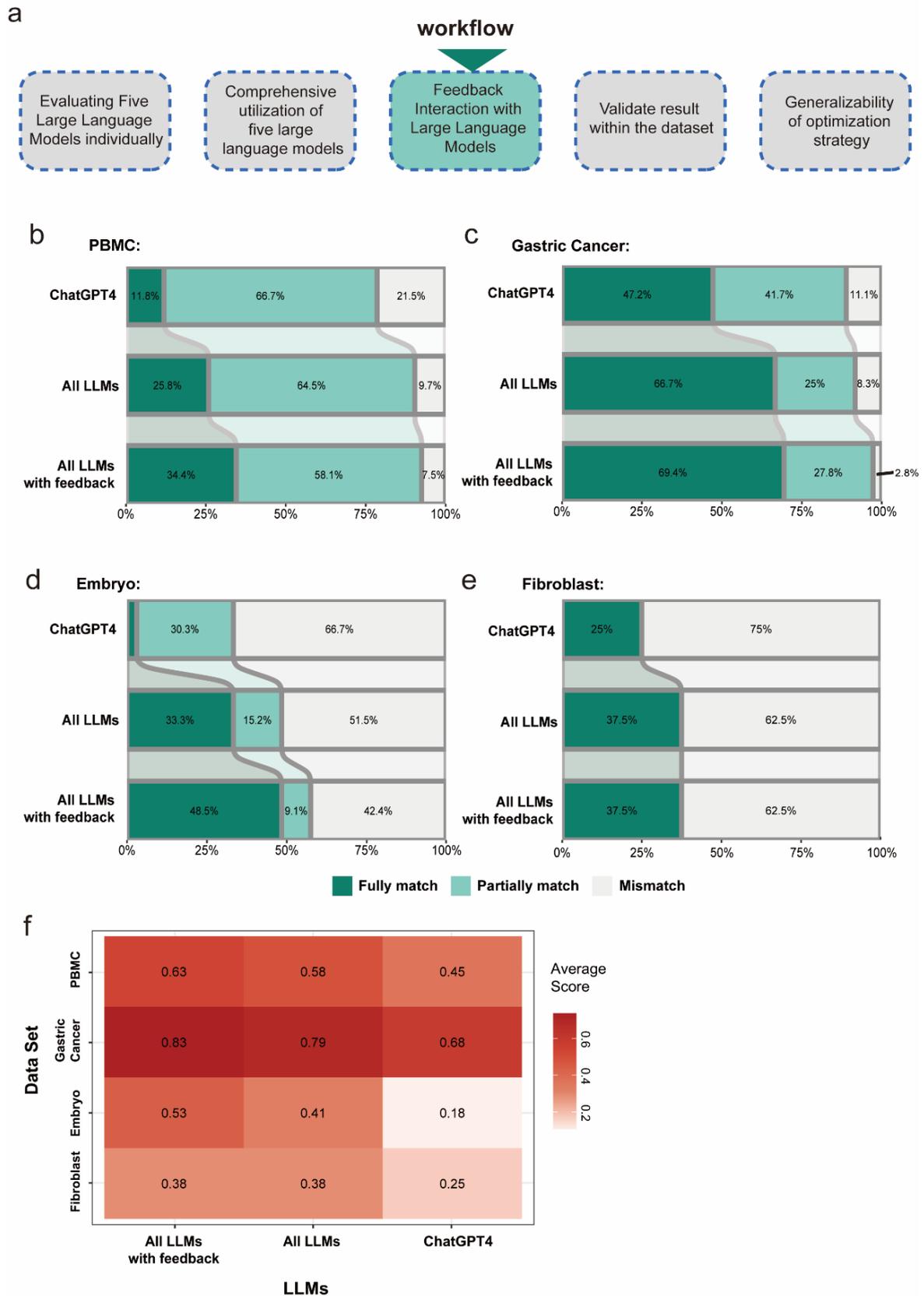

**Figure 5. Reliability of LLM and expert annotations within mismatch results.**

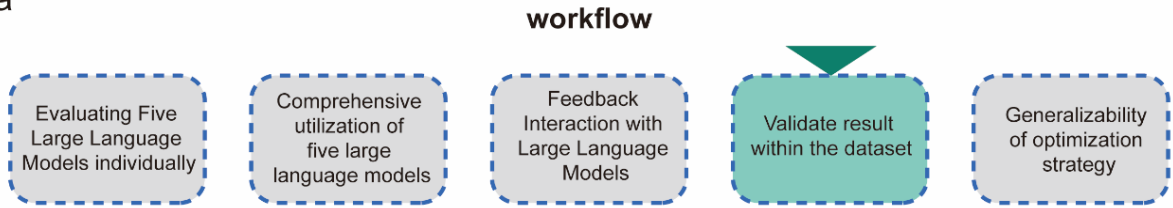

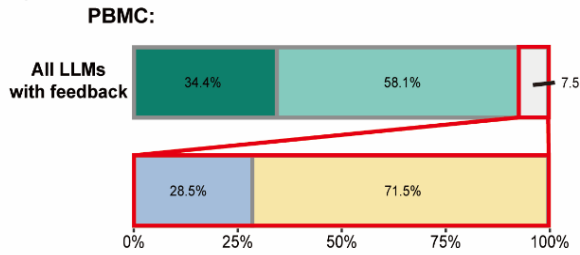

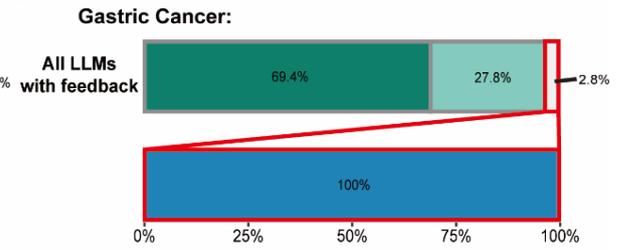

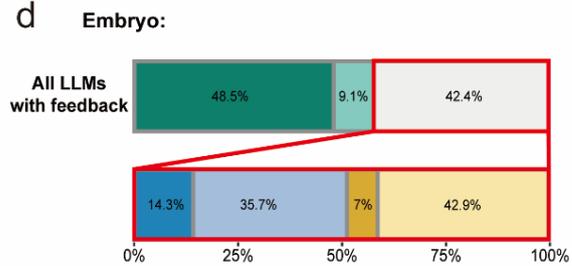

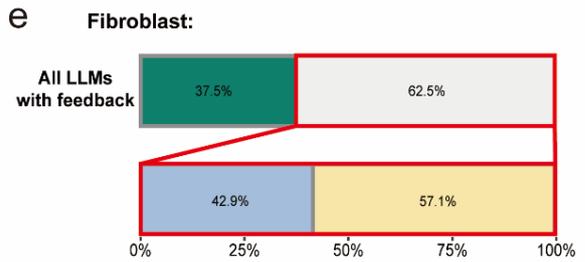

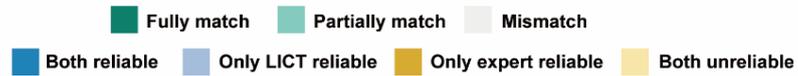

**Figure 6. Generalizability of our optimization strategy.**

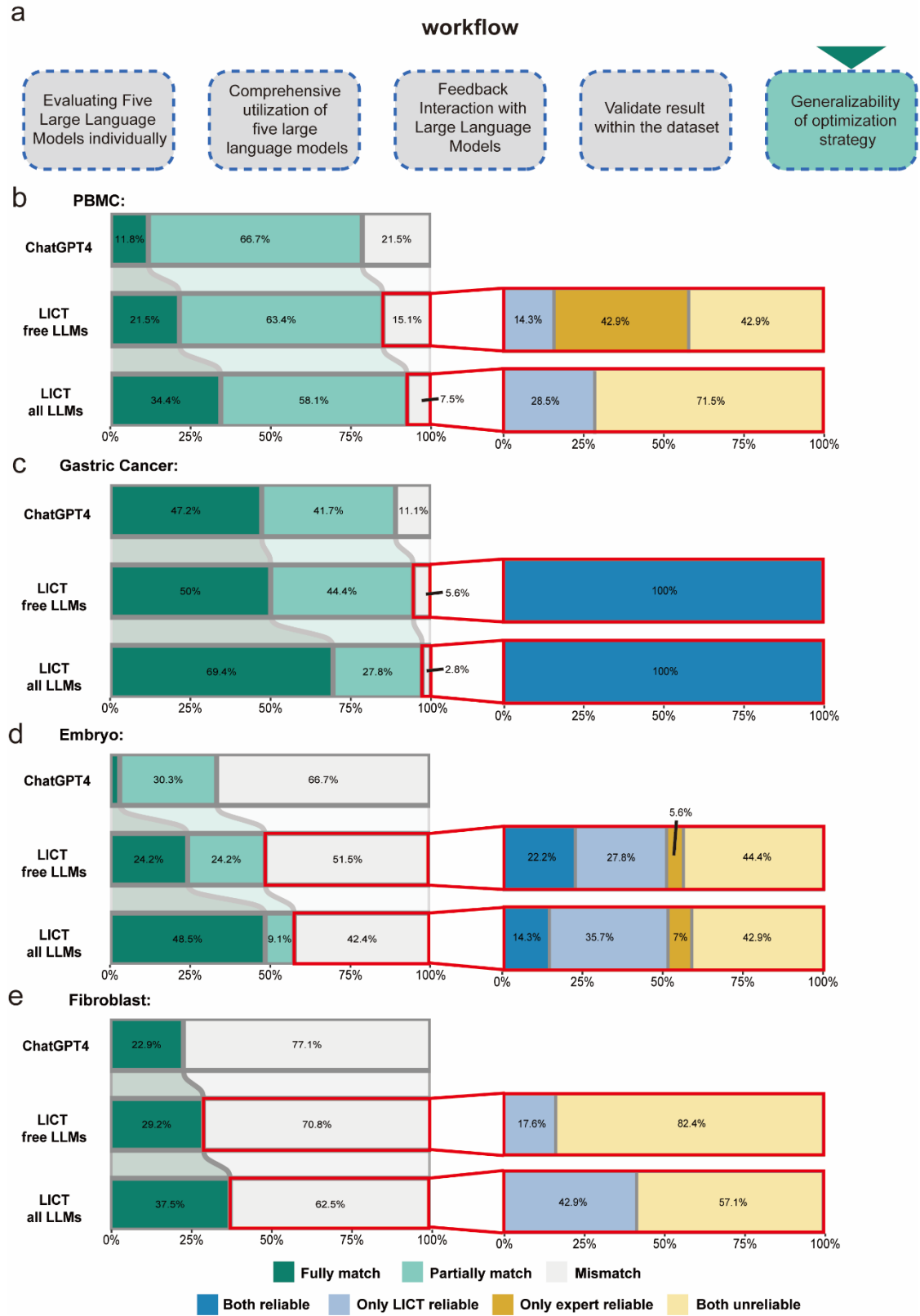

**Table Supplementary1. The large language models aligned with expert annotations**

| Model | Company | Website | Number of cell type | Response number | Free | API | Match | Mismatch |
|---|---|---|---|---|---|---|---|---|
| Claude 3 opus | Anthropic | claude.ai | 31 | 31 | No | Yes | 26 | 5 |
| Llama 3 | Meta | llama.meta.com/ | 31 | 31 | No | No | 25 | 6 |
| ERNIE-4.0 | Baidu | qianfan.cloud.baidu.co | 31 | 31 | No | Yes | 25 | 6 |
| GPT4 | OpenAI | openai.com/ | 31 | 31 | No | Yes | 24 | 7 |
| gemini 1.5 pro | Google | deepmind.google/technologies/gemini/pro/ | 31 | 31 | Yes | Yes | 24 | 7 |
| Mixtral of Expert-8x22b | Mistral | mistral.ai/news/mixtral-8x22b | 31 | 31 | Yes | Yes | 14 | 17 |
| Nemotron-4 340B | NVIDIA | research.nvidia.com/publication/2024-06_nemotron-4-340b | 31 | 26 | N/A | Yes | 12 | 14 |
| BAICHUAN | BAICHUAN AI | www.baichuan-ai.com/home | 31 | 29 | Yes | Yes | 12 | 17 |
| Zidongtaichu | The Chinese Academy of Sciences | taichu-web.ia.ac.cn/ | 31 | 25 | Yes | Yes | 12 | 13 |
| GLM-4 | ZHIPU AI | open.bigmodel.cn/ | 31 | 24 | Yes | Yes | 11 | 13 |
| mootshot 3.5 | Moonshot AI | www.moonshot.cn/ | 31 | 31 | Yes | Yes | 11 | 20 |
| DeepSeek V2 | High-Flyer | www.deepseek.com/zh | 31 | 30 | Yes | Yes | 10 | 20 |
| Gemma 2 | Google | blog.google/technology/developers/google-gemma-2/ | 31 | 25 | N/A | Yes | 10 | 15 |
| GPT3.5 | OpenAI | openai.com/ | 31 | 31 | Yes | Yes | 10 | 21 |
| Yi-large | Beijing Zero One Everything Technology Co., Ltd | www.lingyiwanwu.com/ | 31 | 27 | No | Yes | 10 | 17 |
| ChatGLM | ZHIPU AI | chatglm.cn/ | 31 | 25 | Yes | Yes | 10 | 15 |
| Tongyi Qianwen | Alibaba | tongyi.aliyun.com/qianwen/ | 31 | 31 | Yes | Yes | 10 | 21 |
| Xiaobing | Hong Kong Bombax XiaoIce Company | business.xiaoice.com/ | 31 | 25 | No | No | 10 | 15 |
| YunQue AI | Douyin | m.yunque360.com/product-ai.html | 31 | 29 | Yes | No | 9 | 20 |
| ABAB | Minimax | www.minimaxi.com/ | 31 | 31 | Yes | Yes | 9 | 22 |
| 360GPT_S2_V9 | 360 Security Technology | ai.360.com/ | 31 | 31 | Yes | No | 9 | 22 |
| XuelieHou | Mobvoi Inc. | openapi.moyin.com/ind | 31 | 30 | Yes | Yes | 9 | 21 |
| MathGPT | Beijing Century TAL Education Technology Co., Ltd. | math-gpt.org/?model=unlimited | 31 | 31 | Yes | No | 9 | 22 |
| Yayi | Beijing Wenge Technology Co., Ltd. | www.wenge.com/yayi/ | 31 | 20 | Yes | No | 9 | 11 |

| Name | Company | URL | Col4 | Col5 | Col6 | Col7 | Col8 | Col9 |
|---|---|---|---|---|---|---|---|---|
| Yi-Large | Beijing Zero One Everything Technology Co., Ltd | www.lingyiwanwu.com/ | 31 | 30 | No | Yes | 9 | 21 |
| Reka Flash | Reka | www.reka.ai/news/reka-flash-efficient-and-capable-multimodal-language-models | 31 | 31 | Yes | Yes | 8 | 23 |
| Vicuna | Microsoft | lmsys.org/blog/2023-03-30-vicuna/ | 31 | 16 | Yes | Yes | 7 | 9 |
| Qwen 1.5 | Alibaba Cloud | qwenlm.github.io/zh/blog/qwen1.5/ | 31 | 20 | Yes | Yes | 7 | 13 |
| Yayi | Beijing Wenge Technology Co., Ltd. | www.wenge.com/yayi/ | 31 | 20 | Yes | No | 7 | 13 |
| GPT4-O | OpenAI | openai.com/ | 31 | 20 | Yes | Yes | 6 | 14 |
| Command-R-Plus | Cohere | docs.cohere.com/docs/command-r-plus | 31 | 20 | N/A | Yes | 6 | 14 |
| XuelieHou | Mobvoi Inc. | openapi.moyin.com/ind | 31 | 31 | Yes | Yes | 6 | 25 |
| ZiDongTaiChu | The Chinese Academy of Sciences | taichu-web.ia.ac.cn/ | 31 | 24 | Yes | Yes | 6 | 18 |
| TianGong | Kunlun Tech Co., Ltd. | www.tiangong.cn/ | 31 | 31 | Yes | Yes | 6 | 25 |
| Yi-chat | Beijing Zero One Everything Technology Co., Ltd | huggingface.co/01-ai/Yi-34B | 31 | 31 | No | Yes | 5 | 26 |
| Reka Core | Reka | www.reka.ai/ | 31 | 20 | Yes | Yes | 4 | 16 |
| Gemma 2 | Google | blog.google/technology/developers/google-gemma-2/ | 31 | 20 | N/A | Yes | 4 | 16 |
| Command-R | Cohere | cohere.com/command | 31 | 31 | Yes | Yes | 4 | 27 |
| SparkDesk | Iflytek Co.,Ltd | xinghuo.xfyun.cn/ | 31 | 20 | Yes | Yes | 4 | 16 |
| YunZhiShengShanHai | Unisound AI Technology Co., Ltd. | shanhai.unisound.com/ | 31 | 10 | Yes | Yes | 4 | 6 |
| ChatGLM | ZHIPU AI | chatglm.cn/ | 31 | 20 | Yes | Yes | 4 | 16 |
| Mianbiluka | Mianbizhineng | luca.cn/ | 31 | 31 | Yes | No | 3 | 28 |
| MATHGPT | Beijing Century TAL Education Technology Co., Ltd. | math-gpt.org/?model=unlimited | 31 | 31 | Yes | No | 3 | 28 |
| Openheremes-2.5 | | N/A | 31 | 31 | N/A | N/A | 2 | 29 |
| snowflake-arctic-instrcut | Snowflake | N/A | 31 | 3 | N/A | N/A | 1 | 2 |
| TigerBot | Guangzhou Dongyue Information Technology Co., Ltd | tigerbot.com/login?redirect=/ | 31 | 21 | Yes | Yes | 1 | 20 |

| Name | Company | URL | Col4 | Col5 | Col6 | Col7 | Col8 | Col9 |
|---|---|---|---|---|---|---|---|---|
| XVERSE-Long-256K | XVERSE Technology | chat.xverse.cn/home/index.html | 31 | 26 | Yes | Yes | 1 | 25 |
| DBRX | Databricks | www.databricks.com/ | 31 | 0 | N/A | No | 0 | 0 |
| OLMA | Meta AI | N/A | 31 | 0 | N/A | N/A | 0 | 0 |
| CoVLM2 | ZHIPU AI | http://cogvlm2-online.cogviewai.cn:7861/ | 31 | 0 | N/A | Yes | 0 | 0 |
| Zephyr ORPO | | N/A | 31 | 0 | N/A | Yes | 0 | 0 |
| LLaVA | Microsoft | llava-vl.github.io/ | 31 | 0 | Yes | Yes | 0 | 0 |
| Phi-3-mini-4k-instruct | Microsoft | huggingface.co/microsoft/Phi-3-mini-4k- | 31 | 0 | Yes | Yes | 0 | 0 |
| Vicuna | Baidu | yiyan.baidu.com/ | 31 | 0 | Yes | Yes | 0 | 0 |
| SenseNova | Sense Time | platform.sensenova.cn/home | 31 | 0 | Yes | Yes | 0 | 0 |
| INTERN | Shanghai Artificial Intelligence Laboratory | intern-ai.org.cn/home | 31 | 0 | Yes | Yes | 0 | 0 |
| HunYuan | Tencent | hunyuan.tencent.com/ | 31 | 0 | Yes | Yes | 0 | 0 |
| PanguLM | Huawei | www.huaweicloud.com/product/pangu/nlp.html | 31 | 0 | No | Yes | 0 | 0 |
| Xiaoyi | Huawei | consumer.huawei.com/cn/mobileservices/celia/ | 31 | 0 | Yes | No | 0 | 0 |
| Skywork-MM | Kunlun Tech Co., Ltd. | www.tiangong.cn/ | 31 | 0 | Yes | Yes | 0 | 0 |
| WPS AI | Beijing Kingsoft Office Software, Inc. | www.wps.ai/ | 31 | 0 | Yes | No | 0 | 0 |
| QiYuan | Qihoo 360 Technology Co. Ltd | qiyuan.360.cn/ | 31 | 0 | N/A | No | 0 | 0 |
| Tonghui | Meituan | N/A | 31 | 0 | N/A | No | 0 | 0 |
| Ziyue | Youdao | aicenter.youdao.com/#/home | 31 | 0 | Yes | No | 0 | 0 |
| ZhiHaiTu AI | Beijing Zhizhe world Technology Co., Ltd. | N/A | 31 | 0 | N/A | No | 0 | 0 |
| AntGLM | Ant Group Co., Ltd. | github.com/alipay/PainlessInferenceAcceleratio | 31 | 0 | N/A | No | 0 | 0 |
| BlueLM | vivo Mobile Communication Co.Ltd. | developers.vivo.com/product/ai/bluelm | 31 | 0 | N/A | No | 0 | 0 |
| WangYiYouXiangZhiNengZhuSh | NetEase, Inc. | N/A | 31 | 0 | N/A | No | 0 | 0 |
| XVERSE-Long-256K | XVERSE Technology | chat.xverse.cn/home/index.html | 31 | 20 | Yes | Yes | 0 | 20 |
| 39AIQuanKeYiSheng | Guiyang Longmaster Information and Technology Co., Ltd. | N/A | 31 | 0 | N/A | No | 0 | 0 |
| Infinity Finance | Transwarp Technology (Shanghai) Co., Ltd. | www.transwarp.cn/subproduct/infinity-intelligence | 31 | 0 | Yes | No | 0 | 0 |
| SoulX | Transwarp Technology (Shanghai) Co., Ltd. | N/A | 31 | 0 | N/A | No | 0 | 0 |
| Dialogue Foundation Model | Stepfun | N/A | 31 | 0 | N/A | No | 0 | 0 |
| CVTE | Guangzhou Shiyuan Electronic Technology Company Limited | N/A | 31 | 0 | N/A | Yes | 0 | 0 |
| YunTianTianShu | Shenzhen Intellifusion Technologies Co., Ltd. | N/A | 31 | 0 | N/A | Yes | 0 | 0 |
| Zhejiang-XiHudaMoXing | Scietrain | xinchenai.com/model | 31 | 0 | No | Yes | 0 | 0 |
| HithinkGPT | Hithink RoyalFlush Information Network Co., Ltd | aimiai.com/ | 31 | 0 | Yes | No | 0 | 0 |
| Hairuo | INSPUR GROUP CO.,LTD. | cloud.inspur.com/hairuo/index.html | 31 | 0 | No | Yes | 0 | 0 |
| MiracleVision | Meitu, Inc. | www.miraclevision.com/ | 31 | 0 | N/A | No | 0 | 0 |
| ChatJD | JD.com, Inc. | yanxi.jd.com/ | 31 | 0 | No | No | 0 | 0 |
| FuLuGua | ByteDance | N/A | 31 | 0 | N/A | No | 0 | 0 |
| LingoWhal | DeepLang | www.lingowhale.com/ | 31 | 0 | Yes | Yes | 0 | 0 |
| MengZi GPT | Beijing Langboat Technology | www.langboat.com/portal/mengzi-gpt | 31 | 0 | Yes | Yes | 0 | 0 |
| CharacterGLM | Beijing Lingxin Intelligent Technology Co., Ltd. | www.ai-beings.com/#/platform | 31 | 0 | Yes | Yes | 0 | 0 |
| KwaiYii | KUAISHOU | github.com/kwai/Kwai | 31 | 0 | N/A | No | 0 | 0 |
| XiaoBing | Hong Kong Bombax XiaoIce Company | business.xiaoice.com/ | 31 | 0 | No | No | 0 | 0 |
| bilibili index | Shanghai Wide Entertainment Digital Technology Co., Ltd. | github.com/bilibili/Index-1.9B | 31 | 0 | N/A | No | 0 | 0 |